\documentstyle[12pt]{article}
\newcommand{\be}{\begin{equation}}
\newcommand{\ee}{\end{equation}}
\begin{document}

\centerline{\bf Completely integrable classical systems}
\medskip
\centerline{\bf connected with semi-simple Lie algebras. II.}
\bigskip
\centerline{\bf A. M. Perelomov}
\bigskip\bigskip
\centerline{\bf Abstract}
\medskip\noindent
{\em The complete integrability of a class of dynamical systems of
the type considered in the preceding paper} [1] {\em but with the
potential $V(q)=q^{-2}+cq^2$ is proved.}

\bigskip\noindent
{\sl Note:}\, This is unchanged electronic version of Preprint ITEP -- 27,
Moscow, 1976  
\bigskip\bigskip
\setcounter{equation}{0}

\noindent
Let us consider the classical dynamical system with $n$ degrees of freedom
that is described by the Hamiltonian
\be H(p_k, q_l) = \frac {1}{2}\,\sum _{k=1}^n\,p_k^2+U(q_1,\ldots ,q_n).\ee

The preceding paper by Olshanetsky and Perelomov [1] (see also preceding
papers [2] -- [4]) considers the systems of the type (1)
with a potential
\be
U_0(q_1,\ldots ,q_n) = g^2\,\sum _{k<l}\,[V(q_k-q_l)+\varepsilon
V(q_k+q_l)] +g_1^2\,\sum _{k=1}^n\,V(q_k)+g_2^2\,\sum _{k=1}^n\,V(2q_k),
\ee
where constants $\varepsilon , g, g_1$ and $g_2$ satisfy the condition
\[ \varepsilon =0\,\,\mbox{or\,\,1\,\,for}\,\,g_1=g_2=0;\quad \varepsilon =1,
\,\,\mbox{for}\,\,g_1^2+g_2^2>0
\]
and the potential $V(q)$ has one of the following forms
\begin{description}
\item[I.]   \[ V(q)=q^{-2}, \]

\item[II.]  \[ V(q)=a^2\,\sinh ^{-2}aq, \]

\item[III.] \be V(q)=a^2\,\sin ^{-2}aq, \ee

\item[IV.]  \[ V(q) =a^2\,{\wp }(a q)\]
\end{description}
(${\wp}(q)$ is the Weierstrass function). In  paper [1] it was
also shown that these systems are related to the semi-simple Lie
algebras.

The aim of the present paper is to extend results of paper [1] to
the systems of the type (1) with the potential (2), where
$V(q)=q^{-2}+cq^2$.\footnote{After this paper has been performed
we got the possibility to acquaint with Adler's preprint [6] in
which the similar results were obtained.} Here, besides the
integrals of motion $I_k$, we obtain also the explicit expressions
for quantities $B_k^*$ and $B_k$ which are a classical analog of
quantum operators of creation and annihilation $B_k^+$ and $B_k$
introduced in paper [5]. Note that in paper [5] the algorithm  was
given for finding of  the operators $B_k^+$ and $B_k$, and the
explicit form of the first three of them was
found.
\footnote{\,Note that analogously to paper [4] it may be
shown that in the quantum case ($p_k\to -i\,{\partial }/ {\partial
q_k}$) operators $B_k$ and $B_k^+$ are well-defined. Thereby, we
get for them the explicit expressions.}

Note first of all that under the fulfillment of condition
$\sum _{j=1}^n\, q_j = 0$ the system of interest described by the
Hamiltonian (1) with the potential (2) and $V(q)=q^{-2}+\alpha ^2q^2$
is equivalent to the system with the potential
\be
U=U_0+\frac{\omega ^2}{2}\,\sum _{k=1}^n\,q_k^2,
\ee
where $U_0$ is given by the formula (2) with $V(q)=q^{-2}$.

In order to prove the complete integrability of the system under
consideration and to find explicit form of $B_k, B_k^*$ and $I_k$
we generalize the Moser method [2] which in turn applies
the Lax trick [7]. Namely, we will show that the Hamilton equations
\be
\dot q_k = \frac{\partial H}{\partial p_k}=p_k,\,\,\dot p_k=-\frac
{\partial H}{\partial q_k}
\ee
are equivalent to the matrix equation
\be
\frac{d\tilde L}{dt}=[\tilde L, M]-i\omega \tilde L,
\ee
where matrices $M$ and $L$ depend on $p_k$ and $q_l$, and the matrix $iM$
is Hermitian.

Let us first of all note that as it is shown in papers [1] -- [4]
for the system described by the Hamiltonian
$H=\frac{1}{2}\,\sum _{k=1}^n\,p_k^2+U_0$, where $V(q)$ is given by the
formula (3), the Hamilton equations are equivalent to the matrix equation
\be
\frac{dL}{dt}=[L,M]
\ee
(the explicit form for the matrices $iM=(-D+Y)$ and $L=P+iX$ is given
in the above mentioned papers). In the case of interest we take
\be \tilde L = L-i\,\omega Q,\ee
where the matrix $Q$ is obtained from the matrix $P$ replacing $p_k$ by
$q_k$ and correspondingly the matrix $\tilde L$ is obtained from the matrix
$L$ replacing $p_k$ by $b_k=p_k-i\,\omega q_k$.

Then, as it can be easily checked, equations (6) are a consequence
of the equation (8) and of identities \be \frac{d\tilde L}{dt}
=[L,M]-i\omega (P-i\omega Q),\qquad L=P+[M,Q]. \ee The check of
the first of these identities is very simple while for the check
of the second one it is necessary to make use of explicit
expressions for the matrices $L$ and $M$ from papers \mbox{[1] --
[4]}.

Let us now turn to obtaining of consequences from the equation
(6). Form the quantities \be B_k=\mbox{tr}\,(\tilde
L^k)\,\,\mbox{and}\,\,B_k^*=\mbox{tr}\,\left( \left( \tilde
L^+\right) ^k\right) . \ee Then, as it follows from (6), \be \dot
B_k(t) = -\,ik\omega \,B_k(t),\qquad B_k(t)=\exp \,(-\,ik\omega
t)\, B_k(0). \ee Thus, the quantities $B_k(t)$, not being the
integrals of motion, possess, however, a very simple time
dependence\, \footnote{\,The quantities $B_k(t)$ completely
determine the evolution of considered system. They are rational
functions of variables $p_k$ and $q_l$. Expressing the coordinates
and momenta through $B_k(t)$ and $B_l^*(t)$ we get an explicit
expression for $p_k(t)$ and $q_l(t)$ and thereby completely
integrate the system of Hamilton's equations~(5).}

Let us now find the integrals of motion. It is easy to see that
these are both $B_k^*B_k$ and $\{ B_k,B_k^*\}$. Thus the knowledge
of quantities $B_k$ is sufficient to find the invariants. It will
be more convenient for us to use another set of invariants. Let us
consider matrices $N_1=\tilde L^+ \tilde L$ and $N_2=\tilde
L\tilde L^+$. It follows from (6) that they satisfy the equation
\be
\frac {dN_k}{dt}=[N_k,M],\qquad k=1,2.
\ee
Consequently, the
matrix
\be N=\frac{1}{2}\left( \tilde L^+\tilde L+\tilde L\tilde
L^+\right) =L^2+\omega ^2Q^2
\ee
also satisfies the equation
\be
\frac{dN}{dt}=[N,M].
\ee
Therefore, the eigenvalues of matrix $N$
do not change with time and consequently the quantities
\be
I_k=\mbox{tr}\,(N^k)
\ee
are the integrals of motion.

In conclusion, let us prove that the quantities $B_k$ are in involution,
i.e., $\{B_k, B_l\}=0$. Here
\[ \{A, B\}=\sum _{k=1}^n\,\left( \frac{\partial A}{\partial p_k}\,
\frac{\partial B}{\partial q_k}-\frac{\partial A}{\partial q_k}
\,\frac{\partial B}{\partial p_k}\right )\]
is the Poisson bracket for $A$ and $B$.

Indeed, the quantities $B_k$ are obtained from integrals of motion
$I_k^{(0)}$ for the system described by Hamiltonian $H_0$ replacing in them
$p_k$ by $b_k=p_k-i\,\omega q_k$. With this replacement
\[
\frac{\partial I_k^{(0)}}{\partial p_j}\to \frac{\partial B_k}{\partial b_j},
\qquad \frac{\partial I_k^{(0)}}{\partial q_j}\to \frac{\partial B_k}
{\partial q_j}-i\omega \,\frac {\partial B_k}{\partial b_j}.\]
Thus,
\be \{B_k, B_l\}=\sum _j\,\left( \frac{\partial B_k}{\partial p_j}\,\frac
{\partial B_l}{\partial q_j}-\frac{\partial B_k}{\partial q_j}\,
\frac{\partial B_l}{\partial p_j}\right) = \sum _j\,\left( \frac{\partial B_k}
{\partial b_j}\,\frac {\partial B_l}{\partial q_j}-\frac{\partial B_k}
{\partial q_j}\,\frac{\partial B_l}{\partial b_j}\right) = 0,
\ee
since as it follows from papers [1] -- [4],
\[ \{I_k^{(0)}, I_l^{(0)}\} = 0.\]

\end{document}